\documentclass{article}

\usepackage[utf8]{inputenc}
\usepackage{hyperref}
\usepackage{amsmath}
\usepackage{bm}
\usepackage{amssymb}

\renewcommand{\P}{{\cal P}}
\newcommand{\exv}[1]{{\left\langle{#1}\right\rangle}}
\newcommand{\Ito}{{It$\hat{\mathrm{o}}$}}

\title{Finance from the viewpoint of physics}
\author{A. Jakovac\footnote{e-mail: jakovac@caesar.elte.hu}\\
  Institute of Physics, Eotvos University, H-1117 Budapest, Hungary}
\date{\today}

\begin{document}
\maketitle

\begin{abstract}
    In this note we review the basic mathematical ideas used in finance in the language of modern physics. We focus on discrete time formalism, derive path integral and Green's function formulas for pricing. We also discuss various risk mitigation methods.
\end{abstract}

\section{Introduction}
\label{sec:introduction}

Advanced mathematical methods are used in finance for a long time to understand the functioning of the market. In this continuously fluctuating environment probability theory provides that solid basis, on which the assessment of the present values, and the risk mitigation techniques can be based. This aspect of the market has become even more enhanced after the crisis in 2008. Since then the market is more prudent, collateralization is applied often even for simple products. New, more complicated financial products have appeared, the use of computers in the trading becomes more and more widespread. All of these facts result in the increase of the role of mathematical methods in the finance.

There are numerous well written books on mathematical finance, for example \cite{Hull2006, shreve03,shreve032}. These books, and most of the financial literature uses the phrasing of probability theory that was founded by Kolgomorov \cite{kolmogorov2013} and \Ito{} \cite{Ito1986} in the first half of the XX. century. This approach considers the stochastic process as a measure which can be used for integrating a function (adapted process). This thought nicely fits into the mathematical movements of the early XX. century, namely the raise of measure theory and Lebesque integral.

In the same time, however, a different formalism describing probabilistic processes was also born, mainly driven by physicists, Einstein, Langevin, Fokker, Planck, later Dirac and Feynman. Here we treat the stochastic process as a differential equation (\emph{Langevin-equation}), where in the source term an unusual, fast oscillating function appears, called white noise. The white noise is a normally distributed random function where the correlation between different times is described by a Dirac-delta. In the 1920's, however, it was absolutely unclear how to deal with the Dirac-delta "function". It was only the 1950's where Schwartz gave a mathematically satisfying description \cite{Wikidistribution} as a distribution.

An alternative rephrasing of the Langevin-equations can be given using an integral representation, called \emph{functional (or path) integral}. This approach was initiated by Wiener in the 1920's, but its full weight has obtained by Dirac and Feynman in the 1940's \cite{Wikipathintegral}. With this formulation the same problem appeared as for the Dirac-delta earlier: the continuum limit, except for some elementary cases like the Wiener-integral, seemed to be senseless.

The solution for giving sense for the path integral (and, in fact, for all the quantum field theory) arrived only in the 1970's with the ideas of renormalization (for summary and references c.f. \cite{WikiRG}). The main idea is in fact related to the ones used in defining the Lebesque-integral and the Dirac-delta: we approach the continuum limit through some discretization, and we study the change of the results under the change of the discretization. But, unlike in the case of integrals and the distributions, the continuum limit is much more complicated in this case, and we always must keep referring to the discretization scale. Actually, although this could be seem a bug in the line of thought, it leads to new, measurable effects (running coupling constants, trace anomaly) \cite{Collins1984}.

This solution gave a huge impact on the development of statistical physics and quantum field theory, in disciplines where the formalism strongly relies on the path integral. Present day numerical computations of elementary particle physics use mostly path integral methods in some discretization, and no sooner can the continuum limit be achieved than at the end of the computations. In this way, however, precise numerical results could be obtained (c.f. for example \cite{Borsanyi:2016ksw}).

MC simulations are used in various fields nowadays, including finance. In the financial sector the most models are extensions of the Brownian motion, and so Gaussian MC simulations can be applied to simulate the price movements.

The purpose of this note is to give an introduction to finance in the language of physics. Being so, it is the part of an ongoing effort to bring the ideas of physics into finance and vice versa \cite{Mantegna2000, Baaquie:2002tt, Schmidt2004,Kakushadze:2014bea, Jovanovic2017, Baaquie2018}. 

This note is built up as follows. We define the mathematical space that corresponds to the market (Section \ref{sec:spaceoftrade}), then we discuss the value of a portfolio in Section \ref{sec:valueofportfolio}. In Section \ref{sec:statapproach} we look at the market from the point of view of the statistics, and introduce the tools of treating the price changes in a discretized formulation. In Section \ref{sec:continuous} we turn to the possibility of continuous approximation. In the next section (Section \ref{sec:solutionofstochdiff}) we solve some stochastic differential equations. In Section \ref{sec:risks} we discuss risk mitigation techniques applied in the market, and the way how the assumption of risk neutrality leads to the determination of the price of a derivative (Section \ref{sec:pv}). The paper closes with a Summary section (Section \ref{sec:summary}).

\section{The space of trades}
\label{sec:spaceoftrade}

In order to be able to speak about the financial products we have to define an abstract space that represents the trades. To understand the logics we recall that trading traditionally stems from the exchange of properties of different people, families, tribes, or later firms. All tradeable properties will be called asset, let it be direct material goods like vegetables, cattles or tools, or indirect ones as field, workpower or even the life of a person (which is traded for example when somebody enters the army). The assets can have parameters (for example quality, expiration date etc.), then we treat them as different assets.

The property of a trader usually consists of several assets. They can have a house, two horses, five and a half barrel oil and also three and a half cows if two persons have seven cows together. In the property (we will call it a \emph{portfolio}) thus all assets has some quantity. The property or portfolio is thus the list of all the assets with their available quantity. 

The mathematical structure corresponding to this construction is the \emph{vector space}. Let us denote by $A$ that vector space (asset space or portfolio space) where the basis elements are the assets. Although it can be thought to be infinite dimensional (because, for example, the quality forms a continuum), in practice only a finite number of asset types are traded, so we do not loose anything if we think it as a finite dimensional vector space. We mathematically define the portfolio as an element of the asset space
\[\P\in A.\] 

In finance there is a singled out asset that plays a universal role, and this is money. In economics money has various roles, here we just consider one aspect, the universal exchange tool. We use US dollars as numeraire, and denote the corresponding asset by USD. So if we have ten dollars and two dogs, then our portfolio can be described as $\P=10\mathrm{USD}+2\mathrm{dogs}$. Logical.

\subsection{Loans and other promises}

What makes it  more interesting is that not only the actual goods can be traded in a spot exchange, but other ``financial products'' as well. One of the simplest financial product is a loan. This can be money, but other assets can be lended and borrowed, too.

Who has a debt, has, in some sense, a negative property. If we owe three cows then our portfolio could be written as $-3\mathrm{cow}$. But it is not the most adequate notation, and sometimes it can lead to misunderstandings. The reason is that if we have three cows and owe three cows, the above notation would suggest writing $\P=3\mathrm{cow}-3\mathrm{cow}=0$. But it is not true that we have nothing, because we can use the benefits of the cows, for example we can drink their milk.

Thus, somewhat generalizing the concept of the loan, we will speak about general \emph{promises} or liabilities. A debt can be considered as a promise that we will give (back) a certain asset if we are asked for. The loan is the opposite, somebody have promised us a payoff at some time. In fact the actual assets and the promises on actual assets are the main constituents of the more complicated financial products.

Let us denote the promise with $p$, and its argument is the asset that is promised. The loan is a positive promise, because when it is given, one will possess the given asset.  This means that if we have three cows and owe three cows, then our property is
\begin{equation}
  \P = 3\mathrm{cows} - 3p(\mathrm{cows}).
\end{equation}
Now we can not simplify this equation, this means exactly what we want to. $p$ is defined to be a linear map of the asset space
\begin{equation}
    p : A\to A,\qquad p(\alpha a + \beta b) = \alpha p(a)+\beta p(b)
\end{equation}

A promise, since it concerns future events, can have several more parameters, that is why it is worth to denote them as a function. A usual parameter is the maturity or tenor or expiration time, denoting when the promise is due. If we denote the present time as $t=0$, then 
\begin{equation}
  \P = 3\mathrm{cows} - 3p(\mathrm{cows},T),\qquad T=1y
\end{equation}
means that we should deliver 3 cows in one year from now. $T$ can be a time interval, discussed later.

\subsection{Common financial products}

In this language we can describe a lot of financial products. For example a loan with notional $X$ USD, payed back in parts, can be described as
\begin{equation}
  \label{eq:loan}
  \P = X\mathrm{USD} -\sum_{n=1}^Nc_n p(\mathrm{USD},t_n)-X_rp(\mathrm{USD},T),
\end{equation}
where $c_n$ is the interest rate to be paid at time $t_n$ (for example $t_n=n\mathrm{m}$ for monthly payoff), and $X_r$ is the remainder due at expiration time $T$. To determine the value of the parameters $c_n,\,N,\,T$ and $X_r$ at fixed $t_n$, we can use different techniques discussed later. For a fixed rate loan $c_n$ is constant.

Another product is the futures trade when an asset 'a' is agreed to be bought or sold at a given, \emph{strike} price $K$ at maturity time $T$. If we want to buy that asset, called we are in \emph{long position}, then our portfolio consists of
\begin{equation}
  \label{eq:longfutures}
  \P = p(a,T)-Kp(\mathrm{USD},T).
\end{equation}
If we want to sell the asset, called we are in the \emph{short position}, then our portfolio is
\begin{equation}
  \label{eq:shortfutures}
   \P = -p(a,T)+Kp(\mathrm{USD},T).
\end{equation}

Another interesting parameter of the promise can be its optionality. One of the counterparties may have the right not to fulfill or not to exercise their promise. In this case the two parties are not equivalent. We call the one who possesses the optionality to be in the long position, the other counterparty (who ``sells the optionality'') is in the short position, irrespective whether the promise is about to buy or sell something.

A possible notation for the options is to multiply the possible payoffs by a number $\alpha\in\{0,1\}$. When $\alpha=1$, then the promise is fulfilled, otherwise it is denied. It is also important that who has the right to decide the value of $\alpha$, that we indicate as a $\pm$ index: if the index is $+$, then the portfolio owner has the right to set the value of $\alpha$ (i.e. she is in the long position), if the index is $-$ then someone else determines its value (so the portfolio owner is in the short position with respect to the option).

For example if we agreed that trader 'A' has the option to buy a product 'a' at time (or time interval) $T$ for a strike price $K$ from trader 'B' (European option), then their portfolios read
\begin{equation}
  \P_A = \alpha_+(p(a,T)-Kp(\mathrm{USD},T)),\qquad  \P_B = \alpha_-(-p(a,T)+Kp(\mathrm{USD},T)).
\end{equation}

The exercise date can be also optional, in American option it is any value in $[0,T]$, in Bermudan option there are some fixed dates. Similarly as in the previous case, we can denote its optionality by a subscript $\pm$. An American option can be described as
\begin{equation}
    \P_A = \alpha_+(p(a,T_+)-Kp(\mathrm{USD},T_+)),\qquad  \P_B = \alpha_-(-p(a,T_-)+Kp(\mathrm{USD},T_-))
\end{equation}
where $T_+=T_-\in[0,T].$

We note that the strike price can also be a complicated construction, even depending on the price history. For example we can agree that the buyer of the option has the right to sell a given asset at the average price that was achieved in a given time interval (Asian option), or anything more exotic ones.

We also note that, although the choice of $\alpha$ is completely up to the trader in the long position, sensible traders choose $\alpha=1$ if it is beneficial to them. This makes it possible to determine the price of the option, see later.

\section{Value of the portfolio}
\label{sec:valueofportfolio}

By now we can describe what we have currently. In a trade we exchange two (or more) assets. But the question is, how much is a given asset worth? Clearly no one would bargain away his property, but at the same time everybody wants to achieve the highest price possible.

On the other hand there is not an explicit value measure for the goods. In particular because goods may have hidden advantage for somebody, and this person is willing to buy them at a higher price, too. So the only measure for the value of an asset is that for how much is it used to trade. A well informed trader will trade the asset at exactly the price that is adequate at that moment. The lack of information leads to failed trade, or to \emph{arbitrage}, when an asset can be bought from and sold to different parties, realizing a net profit.

If a market is well informed, and there are a lot of vigilant merchants around, then arbitrage can not be hold for a long time. If it was strictly true, then there would be a single price for each asset. But actually it is just an approximation, since nobody knows that value, and so all the trades modify somewhat the price. A momentary excess in demand will raise the price, while a momentary excess of offers will lower it, and this is repeated time and time again. So, if we insist having a definite price, we have to say that the prices \emph{fluctuate}. 

If we sell or buy several assets, then we trade them separately. This means that the price (value of the portfolio) is a linear map from the asset space and time to the real numbers (actually $\bm R_+$). Thus
\begin{equation}
  \begin{array}[t]{lrll}
    S\,:\, & A\times \bm R& \to \bm R_+ &  \qquad\mathrm{linear}\cr
            & (a,t) & \mapsto S(a,t)&\cr
  \end{array}
\end{equation}
gives the price/value of the asset $a$ at a time $t$.

In a fair business neither of the counterparties lose, both of them give or receive the price which corresponds to the assets they trade. If it is a spot bargain, then both parties know the market price, and this serves as a relation point. But if the payoffs happen in the future, one needs a tool to compute the value of the asset at present. This is the \emph{present value}, and this forms the basis of a fair trade.

\subsection{Discounting a risk free zero coupon bond}

The most simple future payoff is the zero coupon bond, which is $p(\mathrm{USD},T)$, i.e. it pays 1USD at a future time $T$ once. We also assume that it is risk free, meaning that we can count on the payoff with hundred percent certainty. For example we may think of a US government bond. Our task is to tell its value at time $t$, which is called \emph{discounting} the value of the payoff.

To tell the present value, we have to compare the investment in a zero coupon bond to a bank deposit in a safe bank. If it would be more advantageous to invest into a bank deposit, then we would short the zero coupon bond now, and put the money in the bank deposit. A time $T$ the bank deposit would have a higher value, and so we could gain money with zero starting capital. If the investment into the zero coupon bond would be more advantageous, we could do the inverse: we borrow money from a bank, and put it into the bond, and realize a net profit at time $T$. To avoid these arbitrage possibilities, the present values of a risk free zero coupon bond and a risk free bank deposit must agree.

But the bank pays interest rate for all the deposits. In the most simple case it is a fixed annual interest rate $r_1$. Technically the paying of the interest happens periodically in each $dt$ time period, with the corresponding interest rate $r_{dt}$. $r_{dt}$ can be determined from the condition that after one year we get $r_1$ rate (assuming $1/dt$ is integer)
\begin{equation}
  (1+r_{dt})^{[1/dt]}=1+r_1\quad\Rightarrow\quad r_{dt}=(1+r_1)^{dt}-1
\end{equation}
In case $dt\to0$ (called \emph{continuous compounding}) we denote $r_{dt}=dt\,r$. Then
\begin{equation}
  (1+r_{dt})^{[t/dt]} = \left(1+\frac{rt\,dt}t\right)^{t/dt}\stackrel{dt\to0}{\longrightarrow} e^{rt}.
\end{equation}
This also means that $r=\ln(1+r_1)$.

If we deposited $X$USD in the bank at time $t$, at a later time $T$ it is worth $Xe^{r(T-t)}$ USD. This should be compared to the case, when we buy a zero coupon bond at time $t$ with maturity $T$. In an arbitrage-free fair business both should have a value of 1USD at time $T$, so we require $Xe^{r(T-t)}=1$. Thus the value of the zero coupon bond at time $t$ is
\begin{equation}
  X = S(p(\mathrm{USD},T),t) = e^{-r(T-t)} = (1+r_1)^{-(T-t)}.
\end{equation}

This formula makes it possible to determine the value of $c$ for a fixed rate loan. The portfolio was given in \eqref{eq:loan}. In a fair business the value of the portfolio is zero at all times. Let us compute it at time zero (present time), when we have
\begin{equation}
  0=S(\P,0)= X - \sum_{n=1}^N c S(p(\mathrm{USD},t_n),0) - X_r S(p(\mathrm{USD},T),0).
\end{equation}
Let us choose $t_n=n\,\Delta t$, $T=(N+1)\Delta t$, and denote the actual interest rate (which is the risk free interest rate plus the spread) by $r$. Then we find
\begin{equation}
  X = c \frac{e^{-r\Delta t}-e^{-rT}}{1-e^{-r\Delta t}} + X_r e^{-rT},
\end{equation}
and, correspondingly,
\begin{equation}
  c = (X- X_r e^{-rT}) \frac{1-e^{-r\Delta t}}{e^{-r\Delta t}-e^{-rT}}.
\end{equation}
Therefore the condition of arbitrage freeness in the absence of risk leads to a definite price for the zero coupon bond, and a definite value of the fixed rate paying.

\subsection{Discounting the price of an asset}

Let us assume that we have a promise that we are given an asset $a$ at time $T$, so our portfolio is $p(a,T)$. What is the value of the portfolio at time $t$?

What we certainly know is that
\begin{equation}
  \label{eq:SpaT}
  S(p(a,T),T)=S(a,T),
\end{equation}
since the promise is fulfilled then, we obtain the asset, and its price is what is determined by the market at that time. We claim that it is true at other times as well, i.e.
\begin{equation}
  S(p(a,T),t)=S(a,t),
\end{equation}
it does not depend on $T$.

The reason is that if $S(p(a,T),t)>S(a,t)$, then we buy the asset now, and, at the same time we sell the promise of delivery at time $T$. Therefore we have now the asset $a$, payed its value ($-S(a,t)\,\mathrm{USD}$), we promised a delivery of $a$ at time $T$ (this is $- p(a,T)$), and we obtained the price for the promise  $S(p(a,T),t)\,\mathrm{USD}$. Our portfolio therefore reads
\begin{equation}
  \P_1(t) = a - S(a,t)\,\mathrm{USD} - p(a,T) + S(p(a,T),t)\,\mathrm{USD}.
\end{equation}
The value of the portfolio is zero at time $t$. Its value at time $T$, if the promise is fulfilled
\begin{equation}
  S(\P_1,T) = S(a,T)- S(p(a,T),T) + (S(p(a,T),t)-S(a,t)) S(\mathrm{USD},T).
\end{equation}
But the first two term cancel each other by equation \eqref{eq:SpaT}, and so what remains is
\begin{equation}
  S(\P_1,T) = (S(p(a,T),t)-S(a,t)) S(\mathrm{USD},T) >0.    
\end{equation}
Therefore we could gain money. If $S(p(a,T),t)<S(a,t)$, then we build a portfolio
\begin{equation}
  \P_2 = -a + S(a,t)\,\mathrm{USD} + p(a,T) - S(p(a,T),t)\,\mathrm{USD},
\end{equation}
for that $S(\P_2,0)=0$ and $S(\P_2,T)= (S(a,t)-S(p(a,T),t)) S(USD,T) >0$ again. To exclude this arbitrage possibility we need to have $S(p(a,T),t)=S(a,t)$, which we wanted to demonstrate.

We remark that the two cases are somewhat different. If the price of the promise is larger than the actual price, we immediately can realize a profit without any original capital. The other case is feasible if we have the asset previously, otherwise we can not realize the $-a$ part of the portfolio. But, if the asset is liquid enough, there are enough assets in the market to forbid this arbitrage. 

Using this result we can give the price of a futures trade. The portfolio of a long position is given by  \eqref{eq:longfutures}, its price is therefore
\begin{equation}
  S(p(a,T)-Kp(\mathrm{USD},T),t) = S(a,t)-Ke^{-r(T-t)}.
\end{equation}

\section{Statistical approach to the market}
\label{sec:statapproach}

In fact the discounting of an asset price is the only one which is independent on the way the market operates. Already the calculation of the discount factor of a fixed payoff depends strongly on the details, in this case on the interest rate. A fair business takes into account the market rates which, however, fluctuate in time. Therefore we should understand, how the market operates, how the prices are determined, why, and how do they fluctuate. This is a very complicated question, and we can just hope that we find a satisfactory approximation.

The first point we have to clarify is the recording of the prices. Although previously we used a continuous time notation, but it is an abstraction, an approximation. In reality all the recordings have a time stamp that is not infinitely fine. There is a smallest time difference that can be resolved, say $d\tau=1\mu$sec (as an upper estimate), thus all trades and prices can be characterized by an integer; in particular the price of asset $a$ at time $t=n d\tau$ will be denoted as $S_{na}$. We will use a fixed $N$ number of assets, then the vector of all prices is $S_n = (S_{n1}, S_{n2},\dots,S_{nN})$. Sometimes we will put a comma between the two indices in order to avoid misunderstanding, for example we will write $S_{n+1,a}$.

When we think about a dynamic model of price changes we must pin down that in a \emph{complete model} the price in the future must depend solely on the information available at the present. In fact, we can not make decisions based on past events if they are forgotten. The only way of remembering the past events is to make notes (eventually in our memory) about them, and then it is an available information in the present. So we may write generally
\begin{equation}
    S_{n+1} = S_{n} + {\cal F}(\mathrm{information\ available\ at\ present}).
\end{equation}
The factors determining the evolution of the price, of course, are numerous. Moreover, for a quantitative prediction we should have known the actual form of the $\cal F$. Thus predicting the price in the future seems to be impossible.

Still, we can benefit from the generic form above. We may divide the information available at present into three parts. The first part are externalities that do not depend on the status of the market: for example the natural events like wheather, new discoveries, inventions, political or military actions. In a a market model we do not want to describe their dynamics, we take them as given processes, and as such these can be taken into account as an explicit time dependence. We may hope that these effects are slow (usually they are, but for example the weather can have significant influence in certain areas also on daily basis).

The second part of the variables describe the market. Among them there are the asset prices, but other market factors can also be present like forward rates. They appear on both sides of the equation, and we denote them unified with $S$.

The third part is again (mainly) independent on the status of the market, but these are fast processes. They consist, for example, of the momentary intentions of the participants of the market. Let us denote them as $\xi_i$, where $i$ runs through some (large) index set. These processes are in principle well defined, they follow their own dynamics, but it is impossible to tell their time dependence from the knowledge of the asset prices. All in all we have the equation
\begin{equation}
\label{eq:deteq}
    S_{n+1} = S_{n} + {\cal F}_n(S_n, \xi_{in}).
\end{equation}

Were the $\xi_i$ absent from the above equation, we could determine $\cal F$ from the observation of price changes in the past, and eventually recalibrate its form from time to time. But it is hopeless to determine the actual form of the $\xi_i$ functions. What helps us in this situation is that they are numerous, and although they are deterministic one-by-one, their net effect is still something that can be described \emph{statistically}. This means that we \emph{assume} a time dependence for them, solve the above equation for all possible time dependences, and finally we average over the results with some weight. We will assume that these variables are normalized in a way that they fluctuate around zero (their mean is treated as a deterministic effect).

\subsection{Linearization}

Using the fact that the $\xi_i$ effects are small one-by-one, we can power expand the $\cal F$ function to first order
\begin{equation}
    S_{n+1} = S_{n} + {\cal F}_n(S_{n},0) + \xi_{in}\frac{\partial{\cal F}_n}{\partial\xi_{in}}\biggr|_{(S_{n},0)}+\dots.
\end{equation}

The last term is a weighted sum of the $\xi_i$ variables at time index $n$. Now we can argue that the distribution of the sum of mostly independent random variables (with bounded variance) is a \emph{Gaussian}. This is the central limit theorem, and in fact we need to fulfil some conditions that we tacitly assume that is in fact the case here. Thus the last term can be substituted by a single term with some generic coefficient:
\begin{equation}
    S_{n+1} = S_{n} + {\cal F}_n(S_{n},0) + Z_n(S_{n}) \xi_n,
\end{equation}
where the $\xi_n$ variables are all Gaussian distributed random variables with zero mean and unit variance. We will assume that these random variables are independent for different times: indeed, we can argue that there are different trades throughout the world at random times, and so their interrelation is weak. But we must know that this is again an approximation, because if we do not observe all effects, the effective dynamics of the rest will contain memory effects. What we assume is that these memory effects are small.

Although all the formulae are supposed to be written for multi-component variables, it may be useful to write out the indices explicitly. In the multi-component notation the above equation can be written as
\begin{equation}
    \label{eq:multivariateS0}
    S_{n+1,a} = S_{na} + {\cal F}_{na}(S_{n},0) + Z_{na}(S_{n}) \xi_{na}.
\end{equation}
The $\xi_{na}$ random variables are not necessarily independent for different assets
\begin{equation}
    \label{eq:covariancematrix}
    \bm{E}\left(\xi_{na}\xi_{mb}\right) = C_{n,ab} \delta_{nm},
\end{equation}
and so the covariance matrix of the complete noise term reads
\begin{equation}
    \label{eq:covariancematrixtotal}
    \bm{E}Z_{na}\xi_{na}Z_{mb}\xi_{mb} = \delta_{nm} Z_{na}Z_{nb} C_{n,ab}.
\end{equation}
To simplify the treatment, we diagonalize the correlation matrix (which is a symmetric regular real matrix) as
\begin{equation}
    C_{n,ab} = \sum_{k=1}^N \lambda_{nk} v_a^{(nk)}v_b^{(nk)},
\end{equation}
where the $\bm v^{(nk)}$ vectors are eigenvectors of the covariance matrix $\bm C_n$, and they are orthonormal: $\bm v^{(nk)}\bm v^{(n\ell)}=\delta_{k\ell}$. Then we can write \eqref{eq:multivariateS0} as
\begin{equation}
    \label{eq:multivariateS}
    S_{n+1,a} = S_{na} + {\cal F}_{na}(S_{n},0) + \sum_{k=1}^N Z_{n,ak}(S_{n}) \xi_{nk}
\end{equation}
with the volatility matrix
\begin{equation}
    Z_{n,ak} = Z_{na} \sqrt{\lambda_{nk}} v_a^{(nk)},
\end{equation}
and uncorrelated noise terms
\begin{equation}
    \bm{E}\left(\xi_{nk}\xi_{m\ell}\right) = \delta_{k\ell} \delta_{nm}.
\end{equation}
Indeed, the correlation of the noise term reads now as
\begin{equation}
    \bm E\left(\sum_{k=1}^N Z_{n,ak}\xi_{nk}\right)\left(\sum_{\ell=1}^N Z_{m,b\ell}\xi_{m\ell}\right) = \delta_{nm} Z_{na}Z_{nb} \sum_{k=1}^N\lambda_{nk} v_a^{(k)} v_b^{(k)}
\end{equation}
which is exactly the complete covariance matrix \eqref{eq:covariancematrixtotal}.

All the above means that it is enough to have as many random Gaussian variables, as the number of the assets on the market (originally we had much more). These variables can be thought to be independent, and appear in the evolution equations multiplied by the volatility matrix $Z_{n,ak}$. Thus the cumulative distribution of the random variables is
\begin{equation}
    {\cal P}(\{\xi\}) = \prod_{nk}{\cal P}_G(\xi_{nk}),\qquad  {\cal P}_G(\xi) = \frac1{\sqrt{2\pi}}e^{-\frac{\xi^2}2}.
\end{equation}
From now on we suppress the multidimensional indices, treat $Z$ as a matrix $Z_{n,ak}$, and $\xi$ as a vector $\xi_{nk}$.

\subsection{Scaling under changing of the discretization time}

In the above discussion the value of $d\tau$ could be chosen arbitrarily. Our first guess was $1\mu$sec, but just as well could it be $2\mu$sec or even $0.5\mu$sec. What effect does it have on the form of the dynamic equation?

Let us first assume that we want to work with $dt=2d\tau$. This can be thought that we want to tell $S_{n+2}$ from $S_{n}$. When we recursively substitute the equation of $S_{n+1}$ into the equation of $S_{n+2}$ we have a lengthy expression. But the price changes are so very little in this time interval that in the argument of $\mu$ and $\sigma$ functions we can use the previous value. This simplifies the discussion to
\begin{equation}
    S_{n+2} = S_{n} + 2{\cal F}_n(S_{n},0) + \sqrt{2}Z_n(S_{n})\frac{\xi_n+\xi_{n+1}}{\sqrt2},
\end{equation}
where in the last expression we divided and multiplied by $\sqrt{2}$. The distribution of the sum of independent Gaussian random variables is a Gaussian random variable. The correlation matrix coming from the the last expression is thus
\begin{equation}
    \frac12\bm{E}(\xi_{na}+\xi_{n+1,a})(\xi_{n,b}+\xi_{n+1,b}) = \frac12\bm{E}(\xi_{na}\xi_{nb})+\frac12\bm{E}(\xi_{n+1,a}\xi_{n+1,b}) = \delta_{ab}
\end{equation}
is the same as for $\xi_n$. Thus we may write
\begin{equation}
    S_{n+2} = S_{n} + 2{\cal F}_n(S_{n},0) + \sqrt{2}Z_n(S_{n})\xi_n.
\end{equation}
This can be generalized to arbitrary $dt$ (as far the change of the prices in this time interval is negligible): the first term is multiplied by $dt/d\tau$, the second term, on the other hand, by $\sqrt{dt/d\tau}$. 
\begin{equation}
    S_{n+dt/d\tau} = S_{n} + \frac{dt}{d\tau}{\cal F}_n(S_{n},0) + \sqrt{\frac{dt}{d\tau}}Z_n(S_{n})\xi_n.
\end{equation}
We may introduce the notations
\begin{equation}
    \mu_n(S_{n})=\frac1{d\tau}{\cal F}_n(S_{n},0),\qquad
    \sigma_n(S_{n}) = \frac1{\sqrt{d\tau}}Z_n(S_{n}),\qquad
    dS_{n}=S_{n+dt/d\tau}-S_{n},
\end{equation}
and then we can write for the $dt$ discretization time
\begin{equation}
    \label{eq:dSdisc}
    dS_{n} = \mu_n(S_{n})\,dt  + \sigma_n(S_{n})\sqrt{dt}\,\xi_n.
\end{equation}
We remark that in the multi-dimensional case $\sigma$ is a matrix in the asset price space.

This form shows that the continuous time limit is not trivial: not all the variables scale like $dt$, and so in the $dt\to0$ limit the above equation does not go to a differential equation. Indeed, the continuous limit is known as a \emph{stochastic differential equation}.

\subsection{Numerical computation of an expectation value}

In the practical point of view the treatment of \eqref{eq:dSdisc} looks like the following. First we find the solution $S_{n+1}$ depending on the time series $\xi = \{\xi_0,\xi_1,\dots,\xi_n\}$ and on the initial condition $S_0$. Let us denote it
\begin{equation}
    S_{n+1}(S_0,\xi).
\end{equation}
Here we have used the fact that $S_{n+1}$ can depend only on the past events. Then we should calculate the expected value of any function of $S_{n+1}$ by averaging over the possible $\xi$ series over independent Gaussian distributions. In formula this reads
\begin{equation}
    \label{eq:avr}
    \bm E f(S_{n+1}) = \int\limits_{-\infty}^\infty \frac{d^N\xi_0}{(2\pi)^{N/2}}e^{-\frac12\xi^2_0}\dots \frac{d^N\xi_n}{(2\pi)^{N/2}}e^{-\frac12\xi^2_n}  f(S_{n+1}(S_0,\xi)).
\end{equation}
The two equations \eqref{eq:dSdisc} and \eqref{eq:avr} provide a well defined numerical framework to solve any stochastic problem numerically.

Often we use a momentum generation function that is defined as
\begin{equation}
    \label{eq:MGF}
    \bm E e^{JS} = \int\limits_{-\infty}^\infty \frac{d^N\xi_0}{(2\pi)^{N/2}}e^{-\frac12\xi^2_0+J_0S_0}\dots \frac{d^N\xi_n}{(2\pi)^{N/2}}e^{-\frac12\xi^2_n+J_nS_n},
\end{equation}
where $JS=\sum_{a,n} J_{na} S_{na}$ and the $S$ series satisfy \eqref{eq:dSdisc}.

\subsection{Change of variables}

A very interesting consequence of the different scaling properties of the various terms in \eqref{eq:dSdisc} is that, in case of a variable change, a nontrivial factor appears.

Let us assume that we have a new variable $X=f(t,S)$, where $f$ is a smooth function. In discretized case it reads $X_n=f_n(S_n)$. Let us consider the change in $X$ up to ${\cal O}(dt^{3/2})$:
\begin{equation}
    dX_n = f_{n+1}(S_{n+1})-f_n(S_n) = \partial_t f_n(S_n)dt + f_n(S_n + dS_n) - f_n(S_n).
\end{equation}
If all terms were scale as $dt$ in $dS$, then we could power expand $f$ to first order. But the different terms scale in different ways, so we must go until the second order term: 
\begin{equation}
    dX_n = \partial_t f_n(S_n) dt+ \partial_S f_n(S_n) dS_n +\frac12 \partial_S^2 f_n(S_n) dS_n^2+{\cal O}(dS_n^3).
\end{equation}
Here we can use \eqref{eq:dSdisc} for the value of $dS_n$. We remark that in $dS_n^2$ there is a single term that is proportional to $dt$, all other terms are of higher order. Thus we shall write
\begin{equation}
    dX = \partial_t f dt + \partial_S f \left(\mu dt  + \sigma\sqrt{dt}\,\xi\right) +\frac12\sigma^2 \partial_S^2 f dt \xi^2+{\cal O}(dt^{3/2}),
\end{equation}
where we omitted the arguments for brevity (note that $f(t,S)$ is a differentiable funciton, so it is sensible to speak about $\partial_t f$ even if the time steps are discrete). We rewrite this formula as
\begin{equation}
    dX = \partial_t f dt + \partial_S f \mu dt +\frac12\sigma^2 \partial_S^2 f dt + \partial_S f \sigma\sqrt{dt}\,\bar\xi,
\end{equation}
where we introduced a new random variable having zero mean as
\begin{equation}
    \bar\xi = \xi + \frac{\sigma \partial_S^2 f}{2\partial_S f }\sqrt{dt}(\xi^2-1).
\end{equation}
As we see, the change of $X$ is not Gaussian distributed, so $X$ is not a Brownian motion any more. But the difference from a Brownian motion vanishes like $\sim \sqrt{dt}$ as $dt\to0$. So in the limit we can omit the difference of $\bar\xi$ and $\xi$. Then we find
\begin{equation}
    dX = \left(\partial_t f+ \mu \partial_S f  +\frac12\sigma^2 \partial_S^2 f\right) dt + \sigma \partial_Sf \sqrt{dt}\,\xi .
\end{equation}
This is the \emph{\Ito{}-formula}. In case of any number of correlated assets it reads
\begin{equation}
    dX_c = \left(\partial_t f_c+ \mu_a \frac{\partial f_c}{\partial S_a}+ \frac12(\sigma^T\sigma)_{ab}\frac{\partial^2 f_c}{\partial S_a\partial S_b}\right) dt + \frac{\partial f_a}{\partial S_a}\sigma_{ab} \xi_b\sqrt{dt}.
\end{equation}

\subsection{Evolution equation of the distribution functions}

Let us assume that we have a statistical information about the price at present, we know its distribution function ${\cal P}_0(S)$. Then what will be the distribution function at later times?

To give a formal definition for the distribution function we realize that for any quantity $g(\xi)$ depending on a real valued random variable the expected value can be written with the help of the Dirac-delta
\begin{equation}
    \bm E_\xi g(\xi) = \int\limits_{-\infty}^\infty dx \, \bm E \delta(\xi-x) g(x).
\end{equation}
The $g(x)$ does not depend on $\xi$, so we can take it out from the scope of the expected value and obtain
\begin{equation}
    \bm E_\xi g(\xi) = \int\limits_{-\infty}^\infty dx \,{\cal P}(x) g(x),
\end{equation}
where ${\cal P}$ is the distribution function
\begin{equation}
    {\cal P}(x) = \bm E_\xi \delta(\xi-x).
\end{equation}

The question we want to answer is that what is the distribution function of the prices at time $t=ndt$ if we know the distribution ${\cal P}_m(S)$ at time $t=mdt$. What we have to do is to solve the price motion using the equation \eqref{eq:dSdisc}, starting from some $S=S_m$ initial condition at $t=m$, and assuming given $\xi = \{\xi_m,\dots,\xi_{n-1}\}$. Having obtained a solution $S_n(S_m,\xi)$, finally we have to average over all $\xi$ and $S_m$. 

Then we can write for the expected value of any $f(S_n)$ function
\begin{equation}
    \bm E f(S_n) =\bm E_{S_m}\bm E_\xi f(S_n(S_m,\xi)) =\! \int\limits_{-\infty}^\infty\! dS dS'\bm E_{S_m}\delta(S'-S_m) \bm E_\xi \delta(S-S_n(S',\xi))f(S).
\end{equation}
With the distribution functions we can write this expression as
\begin{equation}
    \bm E f(S_n) = \int\limits_{-\infty}^\infty\! dS dS' {\cal P}_m(S') {\cal P}_{mn}(S',S) f(S),
\end{equation}
where
\begin{equation}
    \label{eq:Pnss}
    {\cal P}_{mn}(S',S) = \bm E_\xi \delta(S-S_n(S',\xi)).
\end{equation}

There are different methods to derive this quantity, here we will use the \Ito{} formula, applied to the expectation value of the $f(S_n)$ function above. First let us fix the initial distribution to
\begin{equation}
    {\cal P}_m(S') \to \delta(S'-S_m),
\end{equation}
then the $S'$ integral disappears. Now we change $n$, and write up the change in the expected value in two ways. At the one hand we have
\begin{equation}
    d(\bm E f(S_n)) = \int\limits_{-\infty}^\infty dS\,d{\cal P}_{mn}(S_m,S)f(S).
\end{equation}
At the other hand from \eqref{eq:dSdisc} we have
\begin{equation}
\begin{split}
    d(\bm E f(S_n)) &= \bm E df(S_n) = \bm E \left(\mu \partial_S f  +\frac12\sigma^2 \partial_S^2 f\right)dt = \\
    &= \int\limits_{-\infty}^\infty dS\,\left(\mu \partial_S f  +\frac12\sigma^2 \partial_S^2 f\right) {\cal P}_{mn}(S_m,S)dt = \\
    &= \int\limits_{-\infty}^\infty dS\,f(S) \left(-\partial_S (\mu {\cal P}) +\frac12\partial_S^2 (\sigma^2 {\cal P}) \right)dt,
\end{split}
\end{equation}
where in the last line we performed partial integration, and omitted the arguments of ${\cal P}$ for brevity. Since the two expressions are equal for any $f$ function, we can conclude
\begin{equation}
    \label{eq:dPn}
    d{\cal P} = \left(-\partial_S (\mu {\cal P}) +\frac12\partial_S^2 (\sigma^2 {\cal P}) \right)dt.
\end{equation}
In continuous time this leads to a partial differential equation known as the \emph{Fokker-Planck-equation} or \emph{Kolmogorov-PDE}:
\begin{equation}
    \label{eq:FokkerPlanck}
    \partial_t{\cal P} = -\partial_S (\mu {\cal P}) +\frac12\partial_S^2 (\sigma^2 {\cal P}).
\end{equation}
If we wanted to write out the indices explicitly we would write
\begin{equation}
    \frac\partial{\partial t}{\cal P}_a = -\frac\partial{\partial S_b} (\mu_b {\cal P}_a) +\frac12\frac{\partial^2}{\partial S_b\partial S_c} ((\sigma^T\sigma)_{bc} {\cal P}_a).
\end{equation}

\subsubsection{Composition rule and dependence on the initial conditions}

We can perform the evaluation of the expected value \eqref{eq:Pnss} in two parts, if we want. We choose a $m<k<n$ internal time, and we draw up the condition that at $k$ we arrived at $S=S_k$, and then, starting from this value, we proceed from $k\to n$. Formally we can write
\begin{equation}
    {\cal P}_{mn}(S',S) = \int\limits_{-\infty}^\infty dS'' \bm E_\xi \delta(S''-S_k(S',\xi))\delta(S-S_n(S'',\xi)), 
\end{equation}
where in the last delta function we tacitly assumed that we start the time evolution from $k$. The two Dirac-deltas are independent on each other, because in the first case we have to average only over $\{\xi_m,\dots,\xi_{k-1}\}$, in the second case only over $\{\xi_k,\dots,\xi_{n-1}\}$. Therefore we can write
\begin{equation}
    {\cal P}_{mn}(S',S) = \int\limits_{-\infty}^\infty dS'' {\cal P}_{mk}(S',S''){\cal P}_{kn}(S'',S).
\end{equation}

This formula makes it possible to find a differential equation with respect to the initial conditions of the distribution function. If we change $k$, namely, the left hand side does not vary. Thus
\begin{equation}
    0 = \int\limits_{-\infty}^\infty dS'' \left[d_k{\cal P}_{mk}(S',S'') \right] {\cal P}_{kn}(S'',S) + {\cal P}_{mk}(S',S'')d_k{\cal P}_{kn}(S'',S).
\end{equation}
We can use \eqref{eq:dPn} to write for the first term
\begin{equation}
    \int\limits_{-\infty}^\infty dS'' \left(-\partial_{S''} (\mu {\cal P}_{mk}(S',S'')) +\frac12\partial_{S''}^2 (\sigma^2 {\cal P}_{mk}(S',S'')) \right)dt_k {\cal P}_{kn}(S'',S).
\end{equation}
We perform partial integration, and substitute the result back into the previous expression. Since this must be true for any ${\cal P}_{mk}(S',S'')$ we conclude
\begin{equation}
    d{\cal P}_{kn}(S'',S) = \left(-\mu(S'') \partial_{S''}{\cal P}_{kn}(S'',S) -\frac12\sigma^2(S'')\partial_{S''}^2 {\cal P}_{kn}(S'',S) \right)dt_k.
\end{equation}
In continuous time it reads
\begin{equation}
    \label{eq:FokkerPlanck1}
    \partial_{t_0}{\cal P} = -\mu \partial_{S_0} {\cal P} -\frac12 \sigma^2\partial_{S_0}^2  {\cal P},
\end{equation}
where the $0$ index denotes the initial conditions.

\subsubsection{Change of variables in the distribution function}

We may also work out the change of the distribution function under the change of its argument. We change the variable from $x\to y=Y(x)$, when $Y$ is invertible. Then the distribution of $y$ reads
\begin{equation}
    {\cal P}_y(y) = \int dx {\cal P}_x(x) \delta(y-Y(x)).
\end{equation}
Changing to new variable $y'=Y(x)$, the integral measure changes by the Jacobian, and we find
\begin{equation}
    \label{eq:distofnewvariable}
    {\cal P}_y(y) = \left|\frac{\partial Y}{\partial x}\right|^{-1} \!\!\! {\cal P}_x(x)\;\biggr|_{x=Y^{-1}(y)}.
\end{equation}

\subsection{Path integral}

In \eqref{eq:avr} we have seen how to compute an expectation value numerically. Here we continue this line of thought, rewriting that formula.

To treat \eqref{eq:avr} we have to know the additional information of how to determine $S$, i.e. we need the equation \eqref{eq:dSdisc}. We may work out a formula which is self-contained, i.e. it contains both the time evolution as well as the averaging.
The key is that we can represent a recursion through an integral over a Dirac-delta
\begin{equation}
    f(S_{m+1}) = \int \delta\left(S_m+g_m(S_m,\xi_m)-S_{m+1}\right) f(S_{m+1})dS_{m+1},
\end{equation}
where in the present case \eqref{eq:dSdisc} corresponds to $g_m(S,\xi)=\mu_m(S)dt + \sigma_m(S)\sqrt{dt}\xi_m$. This form can be applied for all $m=1,2\dots n$, and obtain
\begin{equation}
    \bm{E} f(S_n) = \int \prod_{m=1}^n \left[e^{-\frac12\xi^2_m} \delta(S_{m}-S_{m-1}-g_{m-1}(S_{m-1},\xi_{m-1})) \right]\frac{ f(S_n)\, {\cal D}\xi{\cal D}S}{(2\pi)^{Nn/2}},
\end{equation}
where  with initial condition $S_0=$given, and we also introduced the notation
\begin{equation}
    {\cal D}h = dh_1dh_2\dots dh_n
\end{equation}
for $h=\xi$ and $S$.

In order to simplify the formulae, and get rid of the disturbing constant factors, we may introduce
\begin{equation}
    \label{eq:exvfsn}
    \exv{f(S_n)} = \int \prod_{m=1}^n \left[e^{-\frac12\xi^2_m} \delta(S_{m}-S_{m-1}-g_{m-1}(S_{m-1},\xi_{m-1})) \right] f(S_n)\, {\cal D}\xi{\cal D}S,
\end{equation}
and then
\begin{equation}
    \bm E f(s_n) = \frac1{\exv{1}} \exv{f(S_n)}.
\end{equation}

We can also introduce the generator functional
\begin{equation}
    Z[S_0;J] = \left.\exv{e^{J S}}\right|_{S_0}
\end{equation}
where we also indicated the initial condition. We usually denote $Z(S_0)=Z[S_0;0]=\exv1$, it is sometimes called partition function in physics. Then
\begin{equation}
    \bm E f(s_n) = \frac1{Z(S_0)} \left.\exv{f(S_n)}\right|_{S_0}.
\end{equation}
In the sequel we will omit all constant factors in all expected values, the division with the corresponding $Z$ will take care of the correct normalization.

We note that the upper limit of the product term in \eqref{eq:exvfsn} can be extended to infinity. The reason is that if the integrand does not depend on the last variable, then the Dirac-delta simply gives one. In this way we can get rid of the last integral unless $f$ depends on it. Finally we have
\begin{equation}
    \exv{f(S_n)} = \int \prod_{m=1}^\infty \left[e^{-\frac12\xi^2_m} \delta(S_{m}-S_{m-1}-g_{m-1}(S_{m-1})) \right] f(S_n)\, {\cal D}\xi{\cal D}S.
\end{equation}

The next step is to integrate over the $\xi_m$ variables. This is not difficult, because the $g_m$ are linear in this variable. The master formula is
\begin{equation}
    \int\limits_{-\infty}^\infty \frac{d^N\xi}{(2\pi)^{N/2}} e^{-\frac12\xi^2} \delta(A-B\xi) = \frac1{\det B} e^{-\frac12 (B^{-1}A)^2}.
\end{equation}
We then obtain, using \eqref{eq:dSdisc}
\begin{equation}
    \exv{f(S_n)} = \int e^{-\frac12 \sum_{m=0}^\infty dt (\dot S_m-\mu_m)C_m^{-1}(\dot S_m-\mu_m)} f(S_n)\,{\cal D}_C S,
\end{equation}
where we denoted
\begin{equation}
    \dot S_m = \frac{S_{m+1}-S_m}{dt},\quad C=\sigma_m^T\sigma_m,\qquad {\cal D}_C S = \frac{dS_1}{\sqrt{\det C_1}}\dots \frac{dS_n}{\sqrt{\det C_n}}\dots.
\end{equation}
Here we also used that $\det\sigma =\sqrt{\det C}$.
This formula has the big advantage that it does not need any supplementary condition, we can calculate the expectation values simply by performing the integrals.

In physical terms the exponent is called the Hamiltonian, or, in other context, the Euclidean Lagrangian. So we can write
\begin{equation}
    L_m =  \frac12 (\dot S_m-\mu_m)C_m^{-1}(\dot S_m-\mu_m),
\end{equation}
then
\begin{equation}
    \label{eq:PIdisc}
    \left.\exv{f(S_n)} \right|_{S_0}= \int \left.e^{-\frac12 \sum_{m=0}^\infty dt L_m} f(S_n)\,{\cal D}_\sigma S\right|_{S_0},
\end{equation}
which is called the \emph{path integral representation} of the expectation value.

The distribution function is the expected value of the Dirac-delta:
\begin{equation}
    {\cal P}(0,S_0;t,S) = \frac1{Z(S_0)} \int e^{-\frac12 \sum_{m=0}^\infty dt L_m}\delta(S_n-S) {\cal D}_\sigma S \biggr|_{S_0}.
\end{equation}

\section{Continuous approaches}
\label{sec:continuous}

In the previous section we used a discrete representation of the stochastic process. Traditionally, however, the continuous description is used in general. In this section we overview some of them.

\subsection{Langevin-equaiton: a differential equation form}

In physics the usual procedure is to write up a formal differential equation
\begin{equation}
    \frac{dS}{dt} = \mu + \sigma \xi,
\end{equation}
known as the \emph{Langevin-equation}; the symbols $\mu$ and $\sigma$ denote general $f(t,S)$ functions, while $\xi(t)$ is a continuous random variable known as a \emph{white noise}.

In order to reproduce the discretized form \eqref{eq:dSdisc} we have to choose the correlation function of these random variables carefully. The correct choice is
\begin{equation}
    \bm{E}\xi^{(a)}(t)\xi^{(b)}(t') = C_{ab}\delta(t-t'),
\end{equation}
where $\delta(t)$ is the Dirac-delta distribution. In this case, namely, by integrating the Langevin-equation from $t$ to $t+dt$ we obtain
\begin{equation}
    dS = \mu dt + \sigma \int\limits_t^{t+dt}\xi(t')dt'.
\end{equation}
We re-introduce $\xi_n$ as
\begin{equation}
    \xi_n = \frac1{\sqrt{dt}}\int\limits_t^{t+dt}\xi(t')dt'.
\end{equation}
The correlation between $\xi_n$ and $\xi_m$ for different $n\neq m$ is zero, and
\begin{equation}
    \bm{E}\xi^{(a)}_n\xi^{(b)}_n=\frac1{dt}\int\limits_t^{t+dt} \left[\bm{E}\xi^{(a)}(t')\xi^{(b)}(t'')\right] dt'dt'' = C_{ab}.
\end{equation}
Thus the "average" of a stochastic variable must be calculated by dividing the square-root of the time interval, not the time interval itself.

\subsection{Ito calculus: measures}

Equation \eqref{eq:dSdisc} can be thought as a relation for measures. Then $dt$ serves as an ordinary Riemann-measure, while the $dW=\{\sqrt{dt}\,\xi_n\,|\,n=0,\dots\infty\}$ set is interpreted as a probability measure, usually referred to as the Brownian motion. We now discuss the one dimensional case with $C=1$.

\subsubsection{Probability theory in nutshell}

This approach needs somewhat more preparation, and we recommend the interested reader to turn to more detailed description; here we just list the very essence of what we need. The point is that we try to generalize the concept of random variable to continuous "indices". In the discrete version one defines the sample space $\Omega$ that consists of elementary events, like an actual series of results of finite number of dice throwing (e.g. $(1,3,3,2,4,5)$). The event space $\cal F$ is the power set of $\Omega$, consisting of all the subsets of it. Under the union operation this is a $\sigma$-algebra. 

A probability measure is first defined as a function $P:\Omega\to[0,1]$, but it can be lifted to $P:{\cal F}\to[0,1]$ with $P(E)=\sum_{\omega\in E}P(\omega)$. $P$ must satisfy $P(\Omega)=1$. A random variable is $X:\Omega\to\bm R$. The expected value of a random variable is defined as
\begin{equation}
    \bm{E}_PX = \sum_{\omega\in\Omega} X(\omega) P(\omega).
\end{equation}

In the continuous case the problem is that the elementary events (also called atoms), forming $\Omega$, all have zero probability. Therefore the probability measure can be defined only on $\cal F$, which is additive for unions of (countable) mutually disjunct subsets of $\Omega$:
\begin{equation}
    P:{\cal F} \to[0,1],\qquad P(\Omega)=1,\qquad P(\cup_{i\in I} A_i) = \sum_{i\in I} P(A_i),
\end{equation}
where $I$ is a countable index set, and $A_i\cap A_j = \{\}$ for $i\neq j$. The $(\Omega,{\cal F},P)$ set is called probability space.

The generalization of the discrete expected value to continuous case is a stochastic integral denoted by
\begin{equation}
    \bm{E}_PX = \int_\Omega X(\omega) dP(\omega).
\end{equation}
This is defined as a limiting procedure. First define the integral if $X$ is a step function, i.e. $X=\sum_{i\in I} x_i \bm{I}_{A_i}$, where $A_i$ are disjoint elements of $\cal F$ and $\bm{I}_{A_i}(\omega)=1$ if $\omega\in A_i$ and 0 otherwise (indicator function). Then
\begin{equation}
    \bm{E}_PX = \int_\Omega X(\omega) dP(\omega) = \sum_{i\in I} x_i P(A_i).
\end{equation}
Then this definition can be extended to any function that can be approached as a limit of step functions.

\subsubsection{The Ito process}

The integral associated to the $dW$ measure is the It$\hat{\mathrm{o}}$ integral. In our approach, fixing the $dt$ time steps, we can integrate a function that is constant during these time steps (i.e. a fine step function, in mathematics it is called a process adapted to the discretization). The result of the integral is then a stochastic variable 
\begin{equation}
   I_T= \int\limits_0^T \Delta(t) dW(t) = \sum_{n\le T/dt} \Delta(n\,dt) \xi_n \sqrt{dt}.
\end{equation}
This sum is also a Gaussian variable with zero mean and the following variance
\begin{equation}
     \bm{E} I_T^2 = \int\limits_0^T \Delta^2(t)dt,
\end{equation}
as it can be seen from the square of the sum.

It is not hard to see that this definition does not depend on the length of the time intervals, just because of the Gaussian nature of the $\xi_n$ variables. So refining the time mesh we can approach the integral of any functions that can be described as a limit of step functions (measurable functions).

The quadratic variance of the integration measure reads
\begin{equation}
    dW\,dW = dt \xi_n\xi_n = dt + dt(\xi_n\xi_n-1) = dt + {\cal O}(dt^{3/2}),
\end{equation}
and the last term vanish when $dt\to0$. This formula makes the basis of It$\hat{\mathrm{o}}$ calculus.

\subsection{Path integral}

There is also a continuous notation for the path integral. The sum in \eqref{eq:PIdisc} multiplied by $dt$ naturally leads to the integral notation
\begin{equation}
    \int\limits_0^\infty L(t) dt = \sum_{m=0}^\infty dt L_m
\end{equation}
with $t=m\,dt$. Then
\begin{equation}
    \label{eq:PIcont}
    \left\langle f(S(t))\right\rangle = \int {\cal D}_\sigma S\, e^{-\int\limits_0^\infty dt L(t)} f(S(t)),
\end{equation}
where
\begin{equation}
     L(t,\dot S, S) = \frac12 (\dot S-\mu)C^{-1}(\dot S-\mu).
\end{equation}

\section{Solutions of some stochastic differential equations}
\label{sec:solutionofstochdiff}

In this section we discuss some stochastic differential equations, and give their distribution functions. We will always start from the initial condition $S(t=0)=S_0$, or ${\cal P}(t=0,S) = \delta(S-S_0)$.

\subsection{The Brownian motion}

The simplest stochastic equation is when the drift and the variance are constant. Then we can diagonalize the covariance matrix, and so we may deal with one dimensional problems. The equation we have to solve, in the discrete notation reads
\begin{equation}
    S_{n+1} = S_n + \mu dt + \sigma \sqrt{dt} \xi_n,
\end{equation}
where $\xi_n$ are independent Gaussian variables with zero mean and unit variance. The solution of the recursion is very simple
\begin{equation}
    S_n = S_0 + \mu n dt + \sigma \sqrt{dt} \sum_{i=0}^{n-1} \xi_i.
\end{equation}
We introduce
\begin{equation}
    \xi = \frac1{\sqrt{n}} \sum_{i=0}^{n-1} \xi_i,
\end{equation}
which is a Gaussian random variable with zero mean and unit variance. So we have
\begin{equation}
    S_n = S_0 + \mu t + \sigma \sqrt{t}\xi,
\end{equation}
where $t=ndt$. Thus the distribution function reads
\begin{equation}
    \label{eq:BMdist}
    {\cal P}_{BM}(t,S) = \frac1{\sqrt{2\pi t \sigma^2}} e^{-\frac{(S-S_0-\mu t)^2}{2t\sigma^2}}.
\end{equation}

\subsection{Geometric Brownian motion (GBM)}

The most prominent feature of the market prices is that it is not important in which unit we measure the prices. We can use any currencies, gold prices or any other asset price as numeraire, the dynamics of the market is the same. Therefore only the relative price changes must be important. The stochastic differential equaiton that describes this property is simplest
\begin{equation}
    \frac{\dot S}S = \mu + \sigma \xi
\end{equation}
in the Langevin notation.

With new variable $X=\ln S/S_0$ with some $S_0$ we obtain, using the \Ito{} formula
\begin{equation}
    \dot X = \mu -\frac12 \sigma^2 + \sigma\xi.
\end{equation}
This is the Brownian motion discussed above. Using this equation it is usual to give the solution of the GBM as
\begin{equation}
    \label{eq:GBMsol}
    S = S_0\exp \left[\left(\mu -\frac12 \sigma^2\right)t + \sigma\sqrt{t} \xi \right].
\end{equation}

The distribution function of $X$ is the one given in \eqref{eq:BMdist}. The formula \eqref{eq:distofnewvariable} gives the distribution function of $S$, using $S'=S$
\begin{equation}
    \label{eq:GBMdist}
    {\cal P}_{GBM}(t,S) = \frac1S\, \frac1{\sqrt{2\pi t \sigma^2}} \exp\left[-\frac1{2t\sigma^2}\left(\ln \frac S{S_0}-(\mu-\frac12\sigma^2) t\right)^2\right],
\end{equation}
this is a lognormal distribution.

\subsection{Vasicek/Hull-White model}

In finance the mean reverting model means that for long terms the random variable fluctuates around a single value. Such model is the following
\begin{equation}
    \dot S = a(b- S) + \sigma \xi.
\end{equation}
Depending on whether the parameters are time dependent or not, do we call this model (extended) Vasicek or Hull-White model. 
Here we solve the model with constant parameters.

Introduce a new variable $S=e^{-at}R+b$, then
\begin{equation}
    \dot S = e^{-at}\dot R -ae^{-at}R  = -ae^{-at}R + \sigma \xi,
\end{equation}
therefore
\begin{equation}
    \dot R = \sigma e^{at} \xi.
\end{equation}
This equation can be solved to $R$ by a simple integral. So we find for the original variable
\begin{equation}
    S= S_0 e^{-at} + b(1-e^{-at}) +\sigma \int\limits_0^t\!ds\, e^{-a(t-s)}\xi(s).
\end{equation}
This describes a Gaussian random variable with mean
\begin{equation}
    \bar\mu = S_0 e^{-at} + b(1-e^{-at}),
\end{equation}
and variance
\begin{equation}
    {\bar\sigma}^2 = \sigma^2 \int\limits_0^t\!dsds'\, e^{-a(t-s)-a(t-s')}\exv{ \xi(s)\xi(s')}= \sigma^2\frac{1-e^{-2at}}{2a}.
\end{equation}
So the distribution is
\begin{equation}
    {\cal P}_{VHW}(t,S) = \frac1{\sqrt{2\pi \bar\sigma^2(t)}} \exp\left[-\frac{(S-S_0-\bar\mu(t))^2}{2\bar\sigma^2(t)}\right].
\end{equation}
As we see, the mean in long terms goes to $b$, the process fluctuates around it with a variance $\sigma^2/(2a)$.

\section{Risk of a portfolio}
\label{sec:risks}

In the previous sections we discussed the general framework of the price dynamics. Now let us think about the evaluation of the present value of an asset.

The most striking question is that if there are two assets with interest rates $r_1>r_2$, then why is not there an arbitrage possibility? Indeed, the portfolio
\begin{equation}
    {\cal P} = S_2a_1 - S_1a_2
\end{equation}
has zero value at $t=0$, but at $t=T$ it is worth
\begin{equation}
    S({\cal P},T) = S_2 S_1(T) - S_1 S_2(T) = \left(e^{r_1T}-e^{r_2T}\right) S_1S_2 >0.
\end{equation}
So it seems that it is worth to realize this portfolio, we gain money from nothing. 

The main point that we did not take into account is the risk. Let us assume for example that $a_2$ is practically risk-free, while $a_1$ has an annual default risk $d$. The average annual rate thus is $0\times d + r_1\times(1-d)=r_1\times(1-d)$. The risk therefore diminishes the rate.

The first problem here is that it is very hard to tell the exact value of $d$ before a real default will occur. We may give vague estimates, but we can easily miss a factor of two or even ten. As a number example consider the case when $a_1$ pays an interest rate 20\%, $a_2$ has a risk-free rate 10\%. If the default risk is 5\% for $a_1$, then the average interest rate is still $14\%$, so $a_1$ is a better investment. But if the default risk is 10\% then the average rate is 8\%, then already $a_2$ takes over. 

But there is another effect. Let us assume that we can borrow money for rate $r<r_2<r_1$, and we want to buy the assets from a loan. To be sure we hold back a relative amount $c$ as a collateral (usually it is demanded by the bank lending the money, too). So if we have a principal of 1USD we can borrow $1/c$USD, and after a year we have
\begin{equation}
    1\,\mathrm{USD}\; \to\; \frac{r_2-r}c\,\mathrm{USD}.
\end{equation}
This is the leverage effect, resulting that the effective rate of a risk-free investment can be raised to very high. In the ideal case when $c\to0$, any small difference between the risk-free and bank loan rate makes the effective rate grow to infinity.

The first lesson here is that if there were a risk-free investment possibility with higher annual rate than another, then this would indeed cause a very high level arbitrage possibility. Therefore the completely risk-free rate is a unique number.

The second remark is that we can leverage, of course, the risky investments, too. But there is a possibility to lose all the money with non negligible probability rate, then we stay back with the debt liability. This means that we must reserve a higher collateral in the risky case, preparing for the worst case. This will easily make the effective rate much lower than the effective rate for a risk-free investment.

So the real question is that how conservatively, how prudently do the banks evaluate and treat the risk. The practice nowadays is that the banks do not tolerate risky investments too well. This has some psychological factors in it, the market could work in different ways. But the present day practice requires the business to be practically risk-free.

\subsection{Risk mitigation by creating indices}

The assets, of course are not risk-free one-by-one, so we must make efforts to get rid of the risk. We can do it by combining assets into a portfolio. There are two main techniques to do this. The first one is to combine \emph{independent} assets into a single portfolio: these are called \emph{indices}. So we consider the portfolio
\begin{equation}
    {\cal P} = \sum_{i=1}^N w_i a_i.
\end{equation}
The value of the portfolio reads:
\begin{equation}
    S_P = \sum_{i=1}^N w_i S_i.
\end{equation}
We will assume that the $a_i$ assets follow the equation
\begin{equation}
    \dot S_i = S_i(\mu_i+\sigma_i\xi_i),
\end{equation}
where we factored out the price itself, and $\exv{\xi_i\xi_j}=\delta_{ij}$. If $w_i$ are independent of the prices of the underlying assets, then $S_P$ satisfies
\begin{equation}
    \dot S_P = \sum_{i=1}^N w_i S_i(\mu_i+\sigma_i\xi_i) = S_P(\bar \mu + \bar\sigma\xi),
\end{equation}
where
\begin{equation}
  \bar \mu = \sum_{i=1}^N w_i x_i \mu_i,\qquad \bar\sigma^2 = \sum_{i=1}^N w_i^2 x_i^2 \sigma_i^2,
\end{equation}
where $x_i=S_i/S_P$.

To diminish the effective risk, we should minimize the above expression by choosing the correct weights with the constraint that we should keep the value of the portfolio fixed, i.e.
\begin{equation}
    1 = \sum_{i=1}^N w_i x_i.
\end{equation}
Then we have to satisfy
\begin{equation}
    \frac\partial{\partial w_i}\sum_{i=1}^N \left(w_i^2x_i^2\sigma_i^2 -\lambda  w_i x_i\right) = 0,
\end{equation}
where $\lambda$ is a Lagrange multiplicator. This results in
\begin{equation}
    w_i =\frac{\lambda}{2x_i\sigma_i^2}.
\end{equation}
The value of the $\lambda$ comes from
\begin{equation}
    1= \sum_i \frac{\lambda}{2\sigma_i^2} \quad\Rightarrow\quad \lambda = \frac1{\sum_i \frac1{2\sigma_i^2}}.
\end{equation}
Putting all together, after some algebra, we find
\begin{equation}
    \frac1{\bar\sigma^2} = \sum_{i=1}^N \frac1{\sigma_i^2}.
\end{equation}
We see that in this way we can not achieve a complete risk-free portfolio, but we can mitigate the risks of the single underlying assets.

While this is simple in theory, practically it is not simple to reliably make an estimate on the $\sigma_i$ values. It is also a question, how many assets do we want to include in the index, how do we treat the default risk, etc. We can make also the optimization in a different way, for example fixing a given risk and optimizing the effective interest rate. This results in the fact that there are various indices in the market that differ in the way we compute the weights.

\subsection{Risk mitigation by hedging}

The other way we can mitigate the risk is that we combine assets in a portfolio that have interdependent risks. In the market there are asset classes where the asset prices depend on each other, so there is a correlation between the risks. The most simple example of this case is when we consider an asset, and a \emph{derivative} of it. A derivative in this sense is an asset that is built exclusively on the other, underlying asset (e.g. option, swap or similar products).

So let us assume that we have a portfolio where the underlying asset is $a$, and we add some derivatives $a_i$ to it. So we have
\begin{equation}
    {\cal P} = \sum_i \alpha_i a_i - \delta a,
\end{equation}
where the weights $\alpha_i$ and $\delta$ are real numbers. The value of the portfolio is
\begin{equation}
    \label{eq:SPval}
    S_P = \sum_i \alpha_i f_i(t,S) - \delta S,
\end{equation}
where we have denoted the value of the derivatives at time $t$ and at spot price $S$ as $f_i(t,S)$. Now we think about these functions as prices that can be obtained by observing the market.

What is somewhat more complicated here compared with the previous case, is that the price of the portfolio may depend non-linearly on the price of the underlying, and so its dynamics must be computed using the \Ito{} lemma. So, if
\begin{equation}
    \dot S = \mu + \sigma \xi,
\end{equation}
where $\mu$ and $\sigma$ can be $S$ dependent, then we have for the complete portfolio
\begin{equation}
    \label{eq:dotSP1}
    \dot S_P = \partial_t S_P + \mu S\partial_S S_P + \frac12 \sigma^2 S^2 \partial^2_S S_P + \sigma S\partial_S S_P \xi.
\end{equation}
This expression is risk-free, if the term containing $\xi$ is zero. This leads to
\begin{equation}
    \partial_S S_P = 0.
\end{equation}
This would mean, however, that $S_P$ does not depend on $S$, put another way, it is not built on the asset $a$. This contradicts our first equation.

So perfect risk-freeness can not be achieved in this way, either. The best we can do is to ensure vanishing derivative at a given price of the underlying, practically at the actual spot price $S=S_0$. Thus we require
\begin{equation}
    0= \partial_S S_P\biggr|_{S_0}.
\end{equation}
It is usual to introduce the $\Delta$ risk of the portfolio by the definition
\begin{equation}
    \Delta_P = \partial_S S_P\biggr|_{S_0}.
\end{equation}
Risk-freeness at the spot price requires that the delta-risk of the portfolio vanishes
\begin{equation}
    \Delta_P=0.
\end{equation}
It is also said that we have a \emph{delta-neutral} portfolio, or that we hedged out the delta risk. 

Using our portfolio we have
\begin{equation}
    \Delta_P = \sum_i\alpha_i \Delta_i -\delta,    
\end{equation}
where
\begin{equation}
    \Delta_i = \partial_S f_i(t,S_0).
\end{equation}
A delta-neutral portfolio can be achieved using one single derivative with $\alpha=1$ and the underlying, by choosing
\begin{equation}
    \label{eq:Deltaformula}
    \delta = \partial_S f(t,S_0).
\end{equation}

\subsubsection{Higher order hedging and the "greeks"}

There are several issues with the hedging strategy described above. One is that we do not really know the relation of the underlying and the derivative prices. We can observe the spot price of the derivative, i.e. $f(t,S_0)$, but to estimate $\partial_S f(t,S)$ we should know it for any other prices as well. This can not be observed directly, thus we need a market model. So, strictly speaking, what we can do is to use the estimated present value $\tilde f(t,S, {\cal M})$ which already depends on the market model ${\cal M}$.

In practice the market model has some parameters, first of all the (estimated) volatility parameter $\sigma_0$ of the underlying asset. But, since no market model is perfect, the actual market can be described only with a non-constant volatility parameter. So, in this sense not just the price, but also the model has fluctuations. Now the complete analysis of the previous subsection can be repeated with the substitution $S\to \sigma_0$. What we obtain is that for a risk-free portfolio we need both
\begin{equation}
    \partial_S S_P\biggr|_{S_0} = \partial_\sigma S_P\biggr|_{S_0} = 0.
\end{equation}
It is usual to introduce the quantity $\kappa$ (kappa; sometimes it is called ${\cal V}$ vega), the analogue of $\Delta$, corresponding to the price change under the changing volatility parameter:
\begin{equation}
    \kappa = \partial_\sigma S_P(t,S_0,\sigma).
\end{equation}
We need that the kappa value of the complete portfolio is zero (\emph{delta-kappa neutral} position).

Another issue is that we can ensure risk-free portfolio only at a single price $S=S_0$. As soon as the price moves, the risk will grow. Practically one always has to fine-tune the portfolio by adjusting the $\Delta$ (and $\kappa$) to the actual price. If, however, $\Delta$ strongly depends on the price of the underlying, then a sudden price change is hard to follow. This motivates the introduction of $\Gamma$ as the derivative of $\Delta$ (the second derivative of the present value of the derivative)
\begin{equation}
    \Gamma_P = \partial_S\Delta_P(S) = \partial_S^2 S_P(t,S_0,\sigma).
\end{equation}
To ensure stability of a portfolio not just the delta, but also the $\Gamma_P$ should be zero (\emph{delta-gamma neutral} position).

We could continue this analysis, and introduce other "greeks" to denote the higher derivatives, c.f. for example \cite{Wikigreeks}, all characterize the sanity of a portfolio. But usually, besides delta-risk, the kappa and/or the gamma is the most important to hedge out. 

For all the greeks, the risk of the portfolio is the weighted sum of the individual assets
\begin{equation}
    \kappa_P = \sum_i\alpha_i \kappa_i,\qquad
    \Gamma_P = \sum_i\alpha_i \Gamma_i,\dots.
\end{equation}

If, for example, we have two derivatives, then we can require
\begin{equation}
    \delta = \alpha_1 \Delta_1 + \alpha_2 \Delta_2
\end{equation}
to hedge out the Delta-risk, and
\begin{equation}
    0 = \alpha_1\kappa_1 +\alpha_2 \kappa_2
\end{equation}
to hedge out the kappa-risk. If we want to hedge out the gamma-risk as well, we need a third derivative. 

If we continuously monitor the different greeks of the portfolio, we see, how sensitive it is for various ways of price changes. The best practice is to keep all the risks in a given narrow range.

\section{Present value and pricing}
\label{sec:pv}

As we have argued, the market requires the investments to be the possibly most risk-free. This also means that single assets are practically never traded one-by-one, only in portfolios where the risks are mitigated.
But all risk-free portfolios must grow with the same rate, otherwise arbitrage would show up. This means that the  rates of the individual assets play no role at all. Being part of a portfolio, all assets must be treated as if they had a common drift factor. In this artificial world, called the risk-neutral world we find for all derivatives (including the underlying asset)
\begin{equation}
    \label{eq:riskneutralf}
    \frac d{dt} \exv{f(t,S)}_{rn} = r \exv{f(t,S)}_{rn}
\end{equation}
where $rn$ stands for "risk-neutral". The rate itself can be a time dependent function, but it can not depend on the single asset prices.

This equation, in fact, is enough to determine the present value of an asset. We can do it in two equivalent ways, one leading to a differential equation, the other an integral formula.

\subsection{Black-Scholes-Merton formula}

In this approach we consider a portfolio built on an underlying and one derivative. Its value is
\begin{equation}
    S_P = f(t,S)-\delta S.
\end{equation}
If it is in the delta-neutral position, then
\begin{equation}
    \delta = \partial_S f(t,S_0).
\end{equation}

Now we express the time derivative of the portfolio in two ways. On the one hand the portfolio is risk free at $S=S_0$, so we require \eqref{eq:riskneutralf} to be hold
\begin{equation}
    \frac d{dt} S_P(t,S_0) = r S_P(t,S_0).
\end{equation}
We find for our portfolio above
\begin{equation}
    \frac d{dt} S_P(t,S_0) = rf(t,S_0) - r S_0 \partial_S f(t, S_0).
\end{equation}

On the other hand, if $\partial_S S_P(t,S_0)=0$, then from  \eqref{eq:dotSP1} we find
\begin{equation}
    \label{eq:rfpdt}
    \frac d{dt} S_P(t,S_0) = \partial_t f(t,S_0) + \frac12 \sigma^2 S_0^2 \partial^2_Sf(t,S_0).
\end{equation}

Putting the two equations together we find
\begin{equation}
    \partial_t f(t,S_0) + \frac12 \sigma^2 S_0^2 \partial^2_Sf(t,S_0) = r \left(f(t,S_0) - \partial_S f(t, S_0) S_0\right).
\end{equation}
Strictly speaking the above equation is valid only at $t$ and $S_0$. But as the best approximation for the risk-free portfolio, we can demand that it holds for other $S$ as well. This leads to the \emph{Black-Scholes-Merton differential equation}
\begin{equation}
    \label{eq:BlackScholes}
    \partial_t f +rS\partial_S f+ \frac12 \sigma^2 S^2 \partial^2_Sf = rf.
\end{equation}

The solution of the Black-Scholes-Merton model requires initial condition in time and boundary conditions in $S$. This latter is usually omitted, the boundaries being in the infinity. The initial condition of time, on the other hand, is set by the promised payoff in the future
\begin{equation}
    f(T,S) = P(S).
\end{equation}
It is then a final condition, not an initial one, and we should evolve the time backwards in order to obtain the derivative price today at $t=t_0$. This will give the present value of the derivative. 

\subsection{Integral formula}

We can use a different route to have an expression from the condition \eqref{eq:riskneutralf}. First we find
\begin{equation}
    \frac d{dt} e^{-\int_{t_0}^t dt'r(t')} \exv{f(t,S)}_{rn} =0.
\end{equation}
This means that the quantity
\begin{equation}
    M(t,S) = e^{-\int_{t_0}^t dt'r(t')} f(t,S)
\end{equation}
is a random variable whose expected value under the risk-neutral measure is time independent (called to be a martingale under the risk-neutral measure). 

At $t=t_0$ present time we know the price of the asset, $S=S_0$, thus the price distribution is $\delta(S-S_0)$, and so so the expected value $\exv{M(t_0,S)} = f(t_0,S_0)$. From time independence of the expected value of $M$ follows
\begin{equation}
    f(t_0,S_0) = e^{-\int_{t_0}^t dt'r(t')} \exv{ f(t,S)}_{t,rn}.
\end{equation}
If we have a promised payoff $P(S)$ at time $t$, then $f(t,S)=P(S)$ (assuming the promise is fulfilled). Therefore
\begin{equation}
    \label{eq:PVexv}
    f(t_0,S_0) = e^{-\int_{t_0}^t dt'r(t')} \exv{P(S)}_{t,rn}.
\end{equation}
This formula does not assume any underlying market model, so it can be used in general. 

If we write the payoff as an integral over Dirac-deltas, we can write
\begin{equation}
    f(t_0,S_0) = e^{-\int_{t_0}^t dt'r(t')} \int\limits_{-\infty}^\infty dS' P(S') \exv{\delta(S-S')}_{t,rn}.
\end{equation}
The last term is the distribution function in the risk-neutral world:
\begin{equation}
    f(t_0,S_0) = e^{-\int_{t_0}^t dt'r(t')} \int\limits_{-\infty}^\infty dS' {\cal P}_{rn}(t_0,S_0;t,S')\, P(S') .
\end{equation}
This last formula shows that the Green's function of the present value determination is
\begin{equation}
    \label{eq:Greendef}
    {\cal G}(t_0,S_0;t,S) = e^{-\int_{t_0}^t dt'r(t')} {\cal P}_{rn}(t_0,S_0; t,S').
\end{equation}
Using \eqref{eq:FokkerPlanck1} we see that, if the underlying follows a Langevin equation, then the Green's function satisfies
\begin{equation}
    \partial_{t_0}{\cal G} = r {\cal G}
    -\mu \partial_{S_0}{\cal G}
    -\frac12 \sigma^2 \partial^2_{S_0}{\cal G},
\end{equation}
which is the Black-Scholes-Merton equation \eqref{eq:BlackScholes}. This shows that $\cal G$ is the Green's function of the Black-Scholes equation, too. It also proves that $f(t_0,S_0)$ satisfies the Black-Scholes equation, so the integral approach is equivalent to the differential equation approach.

Using path integral formula we can write from \eqref{eq:PIcont}
\begin{equation}
    {\cal G}(t_0,S_0;t,S) =\frac1{Z(S_0)} \int{\cal D}_C S\,e^{-\int\limits_{t_0}^\infty dt'L(t')} e^{-\int\limits_{t_0}^t dt' r(t')} \delta(S(t)-S)\bigr|_{S_0},
\end{equation}
where
\begin{equation}
    Z(S_0) = \int{\cal D}_C S\,e^{-\int\limits_{t_0}^\infty dt'L(t')} \bigr|_{S_0}.
\end{equation}

If there are several payoffs, then the linearity of the above equation tells us that the present values simply add up. So we can generalize the computation of a present value to arbitrary, continuously compounded payoffs $p(t,S)$
\begin{equation}
    f(t_0,S_0) = \int\limits_{-\infty}^\infty dt \int\limits_{-\infty}^\infty dS\, {\cal G}(t_0,S_0;t,S) p(t,S).
\end{equation}
A fixed payoff at time $T$ can be the represented as $p(t,x) = \delta(t-T)P(T)$.

\subsection{Option price in the GBM market model}

To see an example we will compute the present value of the European call option in the geometric Brownian motion market model. The promised payoff of the call option reads
\begin{equation}
    p(t,S) = (S-K)^+\delta(t-T),
\end{equation}
where $x^+=x\Theta(x)$. To determine the present value, we use \eqref{eq:PVexv}. It contains an expected value calculation, where the best is to use the explicit solution \eqref{eq:GBMsol}, where we shall use the drift $\mu=r=$const. Then we find, with $\xi\to-\xi$:
\begin{equation}
    f(0,S) = e^{-rt} \int\frac{d\xi}{\sqrt{2\pi}} e^{-\frac12 \xi^2 }\left(S e^{(r-\frac12\sigma^2)t - \sigma \xi\sqrt t}-K\right)^+.
\end{equation}
The condition of positivity is $\xi < d_-$, where 
\begin{equation}
    d_-= \frac1{\sigma\sqrt t}\left(\ln \frac {S}K + (r-\frac12\sigma^2)t\right).
\end{equation}
Thus we have
\begin{equation}
    f(0,S) = \int\limits_{-\infty}^{d_-} \frac{d\xi}{\sqrt{2\pi}} e^{-\frac12 \xi^2 }\left(S e^{-\frac12\sigma^2 t - \sigma \xi\sqrt t}-Ke^{-rt} \right).
\end{equation}
The negative of the exponent in the first term is
\begin{equation}
    \frac12 \xi^2 + \frac12 \sigma^2 t +\sigma\sqrt t \xi = \frac12(\xi+\sigma\sqrt t).
\end{equation}
We can change variable in the first term to $\xi' = \xi +\sigma\sqrt t :\in [-\infty, d_+]$, then the upper limit of the integration is 
\begin{equation}
    d_+ = \frac1{\sigma\sqrt t}\left(\ln \frac {S}K + (r+\frac12\sigma^2)t\right)    
\end{equation}
Then in both terms we can realize the erf function, and we arrive finally at the \emph{Black-Scholes-formula}
\begin{equation}
    f(0,S) = S \Phi(d_+) - Ke^{-rt}\Phi(d_-),
\end{equation}
where
\begin{equation}
    \Phi(x) = \int\limits_{-\infty}^x \frac{d\xi}{\sqrt{2\pi}} e^{-\frac12\xi^2}.
\end{equation}

A different form for it reads
\begin{equation}
    \frac{f(0,S)}{K e^{-rt}} = e^m \Phi(\frac m z +\frac z2) - \Phi(\frac m z - \frac z2),
\end{equation}
where
\begin{equation}
    m = \ln\frac{S}{K e^{-rt}},\qquad z = \sigma\sqrt t.
\end{equation}
$m$ at $t=0$ is sometimes called moneyness, $m=0$, i.e. $K=S$ corresponds to the at-the-money (ATM) trade.

From this form we can also calculate the greeks, for example
\begin{equation}
    \begin{split}
        \Delta & = \partial_S f = \Phi(d_+) + \frac1{\sigma\sqrt t}\left( {\cal N}(d_+)-\frac{Ke^{-rt}}S{\cal N}(d_-)\right) \\
        \kappa & = \partial_\sigma f = S \frac{\partial d_+}{\partial \sigma} {\cal N}(d_+) - Ke^{-rt}\frac{\partial d_-}{\partial \sigma}{\cal N}(d_-),
    \end{split}
\end{equation}
where ${\cal N}$ denotes the normal Gaussian function, and
\begin{equation}
    \frac{\partial d_\pm}{\partial \sigma} = -\frac1{\sigma^2\sqrt t}\left( \ln \frac SK + rt\right) \pm \frac12 \sqrt t. 
\end{equation}

\section{Summary}
\label{sec:summary}

The goal of this note was to summarize the ideas used in the financial practice in the language of physics. We have used the discrete time description of the time evolution which fits best to the philosophy of the renormalization group.

This note is far from being comprehensive, there are a lot of details missing. Also most of the discussed material is known and was written in various books even in more elaborated way. What makes this note somewhat different is that it puts emphasis on topics that are not usual to discuss (such as discrete time formalism or path integral).

\section*{Acknowledgment}

The author gratefully acknowledges useful discussions with K. Cziszter, G. Fath and Z. Foris. This research was supported by the Hungarian Research Fund under the contract K104292.

\bibliography{refs} 

\begin{thebibliography}{10}

\bibitem{Hull2006}
J.~Hull, {\em Options, futures, and other derivatives}.
\newblock Upper Saddle River, NJ [u.a.]: Pearson Prentice Hall, 6. ed., pearson
  internat. ed~ed., 2006.

\bibitem{shreve03}
S.~E. Shreve, {\em Stochastic Calculus for Finance {I}: The Binomial Asset
  Pricing Model: Binomial Asset Pricing Model}.
\newblock New York, NY: Springer-Verlag, 2003.

\bibitem{shreve032}
S.~E. Shreve, {\em Stochastic Calculus for Finance {II}: Continuous-time
  models}.
\newblock New York, NY: Springer-Verlag, 2003.

\bibitem{kolmogorov2013}
A.~N. Kolmogorov, {\em Foundations of the Theory of Probability}.
\newblock New York, NY; Heidelberg: Martino Fine Books (November 6, 2013),
  2013.

\bibitem{Ito1986}
K.~Ito, {\em An Introduction to Probability Theory}.
\newblock Cambridge, United Kingdom: Cambridge University Press; 1 edition
  (January 1, 1986), 1986.

\bibitem{Wikidistribution}
``Distribution (mathematics).''
  \url{https://en.wikipedia.org/wiki/Distribution_(mathematics)}.

\bibitem{Wikipathintegral}
``Path integral formulation.''
  \url{https://en.wikipedia.org/wiki/Path_integral_formulation}.

\bibitem{WikiRG}
``Renormalization group.''
  \url{https://en.wikipedia.org/wiki/Renormalization_group}.

\bibitem{Collins1984}
J.~C. Collins, {\em Renormalization}.
\newblock Cambridge, United Kingdom: Cambridge University Press, 1984.

\bibitem{Borsanyi:2016ksw}
S.~Borsanyi {\em et~al.}, ``{Calculation of the axion mass based on
  high-temperature lattice quantum chromodynamics},'' {\em Nature}, vol.~539,
  no.~7627, pp.~69--71, 2016.

\bibitem{Mantegna2000}
R.~N. Mantegna and H.~E. Stanley, {\em An introduction to econophysics:
  correlations and complexity in finance}.
\newblock Cambridge, United Kingdom: Cambridge University Press), 2000.

\bibitem{Baaquie:2002tt}
B.~E. Baaquie, C.~Coriano, and M.~Srikant, ``{Quantum mechanics, path integrals
  and option pricing: Reducing the complexity of finance},'' in {\em {2nd
  International Workshop on Nonlinear Physics: Theory and Experiment Gallipoli,
  Lecce, Italy, June 27-July 6, 2002}}, 2002.

\bibitem{Schmidt2004}
A.~B. Schmidt, {\em Quantitative Finance for Physicists: An Introduction}.
\newblock Elsevier Inc, Cambridge, MA 02139: Academic Press; 1 edition
  (December 28, 2004), 2004.

\bibitem{Kakushadze:2014bea}
Z.~Kakushadze, ``{Path Integral and Asset Pricing},'' {\em Quantitative
  Finance}, vol.~15, no.~11, pp.~1759--1771, 2015.

\bibitem{Jovanovic2017}
F.~Jovanovic and C.~Schinckus, {\em Econophysics and Financial Economics: An
  Emerging Dialogue}.
\newblock New York, NY 10016, USA: Oxford University Press), 2017.

\bibitem{Baaquie2018}
B.~E. Baaquie, {\em Quantum Field Theory for Economics and Finance}.
\newblock Cambridge, United Kingdom: Cambridge University Press; 1 edition
  (August 31, 2018), 2018.

\bibitem{Wikigreeks}
``Greeks (finance).'' \url{https://en.wikipedia.org/wiki/Greeks_(finance)}.

\end{thebibliography}
\bibliographystyle{ieeetr} 

\end{document}